\documentclass[%
 preprint,
 superscriptaddress,
 twocolumn,
 10pt,
 amsmath,amssymb,
 aps,
 longbibliography
]{revtex4-2}

\usepackage{graphicx}
\usepackage{dcolumn}
\usepackage{bm}
\usepackage{xcolor}
\usepackage[utf8]{inputenc}
\usepackage{amsmath,amsthm,amssymb}

\usepackage{tikz,tkz-tab}{{}}
\usepackage[
colorlinks=true,
urlcolor=blue,
linkcolor=blue,
citecolor=blue,
hypertexnames=false
]{hyperref}
\usepackage[normalem]{ulem}
\usepackage{braket}

\def\urlprefix{}
\def\url#1{}

\newcommand{\bfA}{{\textbf{a}}}
\newcommand{\bfB}{{\textbf{b}}}
\newcommand{\bfC}{{\textbf{c}}}
\newcommand{\bfD}{{\textbf{d}}}
\newcommand{\bfE}{{\textbf{e}}}
\newcommand{\bfF}{{\textbf{f}}}

\newcommand{\vtg}{V_\mathrm{{TG}}}
\newcommand{\vlg}{V_\mathrm{{LG}}}
\newcommand{\vrg}{V_\mathrm{{RG}}}
\newcommand{\vlga}{V_\mathrm{{LG,1}}}
\newcommand{\vrga}{V_\mathrm{{RG,1}}}
\newcommand{\vlgb}{V_\mathrm{{LG,2}}}
\newcommand{\vrgb}{V_\mathrm{{RG,2}}}
\newcommand{\vd}{V_\mathrm{{diff}}}
\newcommand{\vs}{V_\mathrm{{sum}}}
\newcommand{\xfree}{X_\mathrm{{2D}}}

\newcommand{\xwire}{X_\mathrm{{1D}}}
\newcommand{\xdot}{X_\mathrm{{0D}}}
\newcommand{\rg}{R_\mathrm{{G}}}

\renewcommand{\thefigure}{\arabic{figure}}

\begin{document}
	
\title{Electrically defined quantum dots for bosonic excitons}

\author{Deepankur Thureja}
\thanks{These authors contributed equally}
\affiliation{%
 Institute for Quantum Electronics, ETH Zurich, Zurich, Switzerland\\
}%
\affiliation{
 Optical Materials Engineering Laboratory, Department of Mechanical and Process Engineering, ETH Zurich, Zurich, Switzerland
}%
\author{Emre Yazici}
\thanks{These authors contributed equally}
\affiliation{%
 Institute for Quantum Electronics, ETH Zurich, Zurich, Switzerland\\
}%
\author{Tomasz Smole\'nski}
\thanks{These authors contributed equally}
\affiliation{%
 Institute for Quantum Electronics, ETH Zurich, Zurich, Switzerland\\
}%
\author{Martin Kroner}
\affiliation{%
 Institute for Quantum Electronics, ETH Zurich, Zurich, Switzerland\\
}%
\author{David J. Norris}
\affiliation{
 Optical Materials Engineering Laboratory, Department of Mechanical and Process Engineering, ETH Zurich, Zurich, Switzerland
}%
\author{Atac {\.I}mamo{\u{g}}lu}
\email{imamoglu@phys.ethz.ch}
\affiliation{%
 Institute for Quantum Electronics, ETH Zurich, Zurich, Switzerland\\
}%

\maketitle

\textbf{Quantum dots are semiconductor nano-structures where particle motion is confined in all three spatial dimensions. Since their first experimental realization, nanocrystals confining the quanta of polarization waves, termed excitons, have found numerous applications in fields ranging from single photon sources for quantum information processing to commercial displays. A major limitation to further extending the range of potential applications has been the large inhomogeneity in, and lack-of tunability of, exciton energy that is generic to quantum dot materials. Here, we address this challenge  by demonstrating electrically-defined quantum dots for excitons in monolayer semiconductors where the discrete exciton energies can be tuned using applied gate voltages. Resonance fluorescence measurements show strong spectral jumps and blinking of these resonances, verifying their zero-dimensional nature. Our work paves the way  for realizing quantum confined bosonic modes where nonlinear response would arise exclusively from exciton--exciton interactions.} 



\begin{figure*}
	\includegraphics[width=12.1cm]{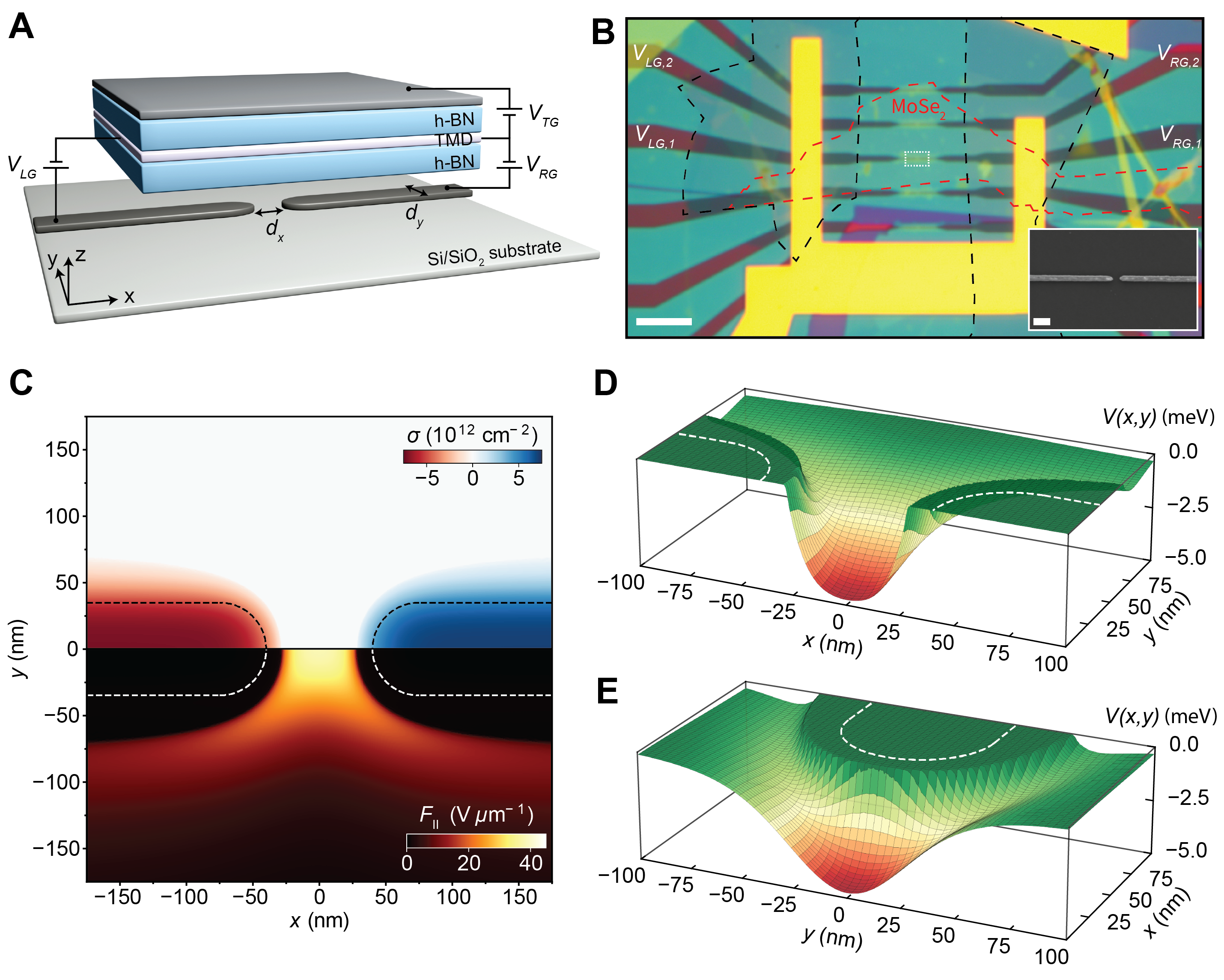}
	\caption{\textbf{Tunable 0D quantum confinement of excitons using split gate electrodes.}
    (\bfA) Proposed device architecture for realizing a 0D quantum confining potential for excitons. The monolayer TMD semiconductor is encapsulated by h-BN. The bottom gate is fabricated on a Si/SiO$_2$ substrate and consists of a pair of split gate electrodes $\vlg$ and $\vrg$, analogous to a quantum point contact, which are separated by a distance $d_x$ and have the width $d_y$. The top gate extends across the entire TMD area. (\bfB) Optical micrograph of the fabricated vdW heterostructure. The red dashed line indicates the monolayer MoSe$_2$ flake, which is encapsulated in h-BN, and deposited on a substrate featuring five pairs of individually controllable finger gates. We investigate the confining potential realized by applying gates voltages to the pair ($\vlga$, $\vrga$) and ($\vlgb$, $\vrgb$), which we term Device 1 and Device 2, respectively. Contacts to the TMD monolayer are formed by few-layer graphene (FLG), indicated in black dashed lines. The entire heterostructure is capped with a global TG, also made of FLG (not shown). The scale bar denotes $5\,\mu$m. The inset shows a SEM image of the split finger gates located in the dashed white rectangle with $d_x\sim80\,$nm and $d_y\sim70$\,nm (scale bar: $200\,$nm). The spatial dependence of the expected itinerant charge density in the semiconducting layer, as determined through electrostatic simulations of Device 1, is shown in the top panel of ({\bfC}), where we assume $\vlg=7.5\,\mathrm{V}=-\vrg$ and $\vtg=0\,$V. The lower panel depicts the resulting magnitude of in-plane electric fields $|F_{\parallel}|$. (\bfD), (\bfE) Magnitude of the corresponding 0D confinement potential $V_\mathrm{conf}(x,y)$ for excitons along the $x$- and $y$-axis, respectively. The dashed line in panels ({\bfC}), ({\bfD}) and ({\bfE}) indicates the spatial extent of the finger gate electrodes.
	} 
	\label{fig:concept}
\end{figure*}

In two-dimensional (2D) semiconductors, such as monolayers of transition metal dichalcogenides (TMDs), the elementary optical excitations are excitons---tightly bound electron--hole pairs~\cite{Wang2018a}. Owing to their fermionic constituents, these quasiparticles can be aptly described as weakly interacting composite bosons with mass $m_\mathrm{X}$~\cite{Combescot2015}. If their spatial motion could be restricted to length scales $\ell$ comparable to or smaller than their lateral size, the Bohr radius $a_\mathrm{X}$, their spectrum becomes anharmonic -- a direct consequence of Pauli's exclusion principle acting upon their fermionic components. Quantum emitters originating from tight confinement of excitons in defects exemplify this phenomenon and have been the subject of intensive research over the past decade~\cite{Aharonovich2016,Turunen2022,Montblanch2023}. Conversely, when the temperature $T$ and confinement satisfy $h /\sqrt{2 m_\mathrm{X} k_B T} \gg \ell \gg a_\mathrm{X}$ (where $h$ and $k_B$ are Planck and Boltzmann constants), the center-of-mass (COM) motion of excitons is fully quantized, leading to the formation of zero-dimensional (0D) bosonic excitons. 

Our work delves into this specific regime of excitonic quantum confinement, examining it through the realization of gate-defined excitonic quantum dots (QDs). In stark contrast to devices that provide tight confinement of excitons, the electrically controlled 0D excitons we study allow for dialing in of the emitter energy at will. However, realization of such a confinement represents an outstanding technological challenge. In particular, despite extensive efforts, e.g., to confine dipolar interlayer excitons~\cite{Hagn1995,Rapaport2005,Gartner2007,Vogele2009,Schinner2013,Butov2017}, the confinement length scales attained in earlier experiments were insufficient to realize true 0D confinement. Conversely, our recent work showcased tunable one-dimensional (1D) quantum confinement of the COM motion of intralayer excitons along a single direction to realize excitonic quantum wires~\cite{Thureja2022}. This achievement was facilitated by the advantageous attributes of van der Waals (vdW) heterostructures. In particular, the tightly bound nature of TMD excitons \cite{Goryca2019} permit the application of large gate-defined in-plane electric fields, with magnitudes exceeding $100\,$V/$\mu$m, over a narrowly confined spatial region of approximately $50$\,nm.

The main underlying principle of our confinement mechanism is the exciton dc Stark shift, which is induced by fringing in-plane electric fields generated along the edge of gate electrodes \cite{Thureja2022,Thureja2023,Heithoff2023}. To adapt this approach from 1D to 0D exciton quantum confinement, we design a device~\cite{Thureja2023} illustrated in Fig.\,\ref{fig:concept}\,{\bfA}. It consists of a TMD monolayer (MoSe$_2$) encapsulated in thin hexagonal boron nitride (h-BN) dielectrics. At the heart of our design are the two bottom split finger gate electrodes analogous to a quantum point contact, which are separated by a distance $d_x$ and have the width $d_y$. Applying substantial voltages $\vlg$ and $\vrg$ of opposite polarity to these finger gates induces heavy electron or hole doping in the narrow region of the TMD layer directly above each gate. This biasing condition results in a pronounced in-plane electric field localized within the gap separating the left and right finger gates. The global top gate (TG) electrode serves to ensure charge neutrality of the TMD layer beyond the immediate vicinity of the finger gates.

\begin{figure*}
	\includegraphics[width=12.1cm]{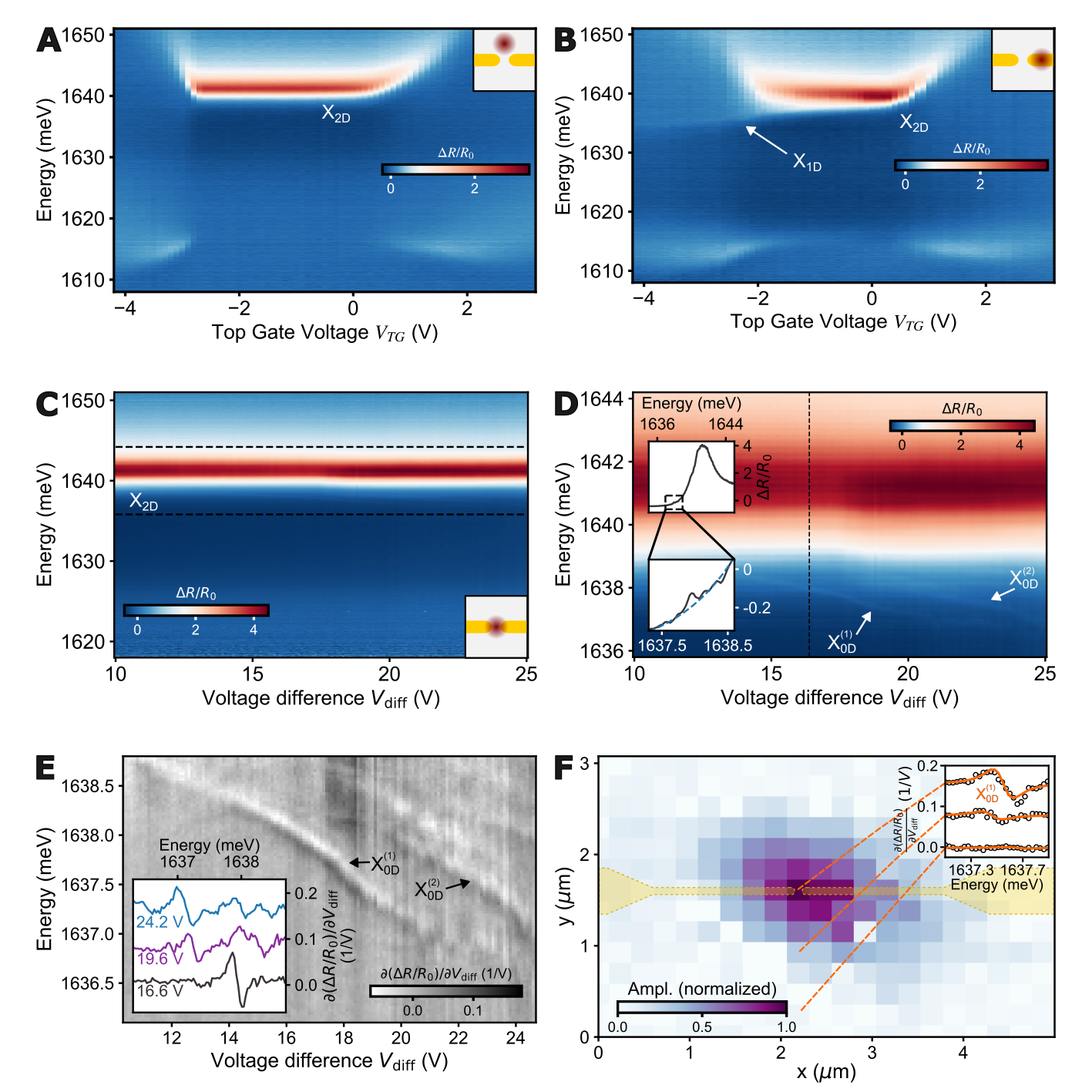}
	\caption{\textbf{Control of exciton dimensionality.}
    Normalized differential reflectance $\Delta R/R_0$ spectra acquired in different regions of the device display signatures of excitons with varying dimensionality. (\bfA) Away from the split gate electrodes the characteristic features associated with a single-gated MoSe$_2$ monolayer are visible. These include the 2D free exciton resonance $\xfree$ for $\vtg$ ranging from approximately $-2.5\,$V to $0\,$V, and repulsive and attractive polaron resonances beyond this range. (\bfB) As the optical spot is shifted to the region $\mathrm{R_R}$, for $\vrg= 12\,$V, an additional narrow redshifting resonance $\xwire$ appears for decreasing $\vtg$, representing 1D exciton confinement perpendicular to the finger gates. (\bfC) Measurement of $\Delta R/R_0$ as a function of $\vd$ at $\vtg = -0.4\,$V over the gap region $\mathrm{R_G}$ reveals a dominant $\xfree$ resonance at an energy around $1641.5$\,meV. On its red tail additional faint redshifting resonances can be observed for increasing $\vd$. Focusing solely on the energy range within the black dashed lines, two resonances, $X_\mathrm{0D}^{(1)}$ and $X_\mathrm{0D}^{(2)}$, can be clearly identified in (\bfD), which we attribute to 0D quantum confined excitons. The inset shows a reflection spectrum line cut at $\vd = 16.6$\,V with a local reflection maximum associated with $X_\mathrm{0D}^{(1)}$ at an energy $E$ around $1638$\,meV. (\bfE) The reflection spectrum derivative $\partial(\Delta R/R_0)/\partial\vd$ exhibits up to five narrow redshifting resonances. Individual spectra for fixed $\vd$ at $16.6\,$V (black), $19.6\,$V (purple) and $24.2\,$V (blue) are shown in the inset. (\bfF) Spatial map of the amplitude of $X_\mathrm{0D}^{(1)}$ at $\vd = 18$\,V, obtained by fitting individual $\partial(\Delta R/R_0)/\partial\vd$ spectra at different locations of the device. The yellow shaded region indicates the location of the finger gates. In the inset, experimental data points are indicated with circles and the corresponding fits are shown with solid orange lines.
    } 
	\label{fig:WL1}
\end{figure*}

Fig.\,\ref{fig:concept}\,{\bfB} presents the optical micrograph of the fabricated device, which comprises five pairs of individually controllable finger gates (see ~\cite{som} for details).
By preparing the finger gate electrodes prior to the device stack we can determine $d_x\sim80\,$nm and $d_y\sim70$\,nm using scanning electron microscopy (SEM) (inset of Fig.\,\ref{fig:concept}). The transferred vdW stack intersects two of these gate pairs, which we designate as Device 1 and Device 2. In the main text, we exclusively focus on measurements performed on Device 1. Additional datasets obtained on Device 2 are presented in \cite{som}.

Such a device configuration provides an effective means to establish a localized p-i-n junction, such that the exciton confinement length scales along the $x$- and $y$-directions, $\ell_x$ and $\ell_y$, can be tuned through appropriate choice of  $d_x$ and  $d_y$, respectively. The spatial dependence of the expected itinerant charge density in the MoSe$_2$ layer, as determined through electrostatic simulations of the fabricated device~\cite{som}, is visualized in the top panel of Fig.\,\ref{fig:concept}\,{\bfC}. We thereby apply voltages of opposite polarity to the finger gates, i.e.~$\vlg=7.5\,\mathrm{V}=-\vrg$, while maintaining $\vtg=0\,$V. The resulting magnitude of the in-plane electric field strength $|F_{\parallel}|$ is illustrated in the lower panel of Fig.\,\ref{fig:concept}\,{\bfC}. Our calculations clearly indicate that a maximum field strength of $|F_{\parallel}^\mathrm{max}| \simeq 40$\,V$/\mu$m can be obtained within a charge-neutral region whose area $A_\mathrm{neutral}$ is determined by the geometry of the finger gates. We emphasize that $|F_{\parallel}^\mathrm{max}|$ can be further enhanced by increasing the difference voltage $\vd = \vlg - \vrg$, which increases the charge densities in the p- and n-doped regions, thus effectively causing a greater potential gradient across a narrower i-region.

The strength of the resulting excitonic COM confinement potential stemming from the dc Stark effect can be determined as $V_\mathrm{conf}(x,y) = -\alpha F_{\parallel}^2/2$, in which $\alpha$ represents the exciton polarizability of MoSe$_2$~\cite{som}. Figs.\,\ref{fig:concept}\,{\bfD} and {\bfE} depict this potential with a cut along the $x$- and $y$-direction, respectively, which suggests the capability of this device architecture to clearly isolate optical signatures associated with 0D excitons from all other background resonances. Firstly, while the charge-neutral global TMD region will inevitably lead to a persistent excitonic resonance, the achievable electric field strengths are sufficient to attain a redshift of approximately $5$\,meV, which is greater than the typical 2D exciton linewidth of $2-3$\,meV at cryogenic temperatures. Furthermore, separation of 0D states from 1D exciton confinement along the finger gate edges is also straightforward, owing to the much shallower depth of the latter when $\vtg\simeq 0\,$V. It is important to note that the complete excitonic confinement potential also includes a repulsive contribution due to exciton--charge interactions \cite{Thureja2022}, which has been deliberately omitted from Fig.\,\ref{fig:concept}\,{\bfD} for the sake of clarity.

To characterize the device and to determine an optimal set of gate voltages which give rise to the doping configuration illustrated in Fig.\,\ref{fig:concept}\,\bfC, we first acquire normalized differential reflectance $\Delta R/R_0$ on a region of the device away from the split gate electrodes. Fig.\,\ref{fig:WL1}\,{\bfA} shows the characteristic doping-dependent behaviour of a single-gated MoSe$_2$ monolayer, where charge neutrality is maintained for $-2.5\,\mathrm{V}\lesssim\vtg\lesssim0\,\mathrm{V}$, as evidenced by a persisting free exciton resonance $\xfree$ in this range. However, dual-gated reflectance experiments, conducted adjacent to a finger gate electrode, suggest a narrower voltage range showing $\xfree$, specifically between $-1\,$V and $0\,$V (Fig.\,\ref{fig:WL1}\,{\bfB}). Within this voltage range, charge neutrality in the extended MoSe$_2$ layer is preserved regardless of the doping state of the regions affected by the left ($\mathrm{R_L}$) and right finger gate electrodes ($\mathrm{R_R}$) \cite{som}. Notably, while the $\vtg$-dependent $\Delta R/R_0$ spectrum displayed in Fig.\,\ref{fig:WL1}\,{\bfB} is dominated by the neutral exciton, as well as the repulsive (RP) and attractive (AP) polaron resonances, a narrow resonance that redshifts with decreasing $\vtg$ emerges for $\vtg \le -1.5\,$V. This $\xwire$ resonance originates from confinement of exciton motion perpendicular to the finger gate, resulting in a 1D exciton mode \cite{Thureja2022,Heithoff2023}.

The influence of the excitonic confining potential generated by the finger gates in the gap region $\mathrm{R_G}$, situated between $\mathrm{R_L}$ and $\mathrm{R_R}$, can be discerned by measuring $\Delta R/R_0$ of Device 1 while keeping the monolayer away from $\mathrm{R_G}$ charge-neutral ($\vtg \simeq -0.4\,$V) and tuning $\vd$, as illustrated in Fig.\,\ref{fig:WL1}\,{\bfC}. Unless stated otherwise, we maintain the voltage sum $\vs = \vlg + \vrg$ at $0\,$V throughout our experimental investigation. Upon positioning the optical spot, which encompasses an area $A_\mathrm{opt}$ of approximately $0.7\,\mu$m$^2$, over $\mathrm{R_G}$, we observe that the $\Delta R/R_0$ spectrum predominantly features the 2D neutral exciton $\xfree$, with an energy near $1641$\,meV. This resonance originates from regions of the MoSe$_2$ monolayer that are covered by the optical spot but are not affected by the narrow finger gates. Intriguingly, when we focus on the red side of $\xfree$, we observe faint resonances that exhibit a pronounced red shift as $\vd$ increases (Fig.\,\ref{fig:WL1}\,{\bfD}); we identify these resonances as $X_\mathrm{0D}^{(1)}$ and $X_\mathrm{0D}^{(2)}$. The insets of Fig.\,\ref{fig:WL1}\,{\bfD} display the reflection spectrum line cut at $\vd = 16.6$\,V, which clearly showcase a local reflection maximum associated with $X_\mathrm{0D}^{(1)}$ at an energy $E$ around $1638$\,meV, situated at the tail of $\xfree$. We tentatively attribute this peak to a QD resonance.


Owing to their small oscillator strength, the weakly confined excitonic states, which are highly sensitive to variations in $\vd$, become more pronounced when the reflection spectrum is differentiated with respect to $\vd$ \cite{som}. Fig.\,\ref{fig:WL1}\,{\bfE} shows the derivative of the reflection spectrum $\partial(\Delta R/R_0)/\partial\vd$, revealing up to five narrow resonances that red shift linearly for increasing $\vd$. The inset of Fig.\,\ref{fig:WL1}\,{\bfE} presents the line shape of these resonances at various $\vd$ values. 

To conclusively show that the resonances depicted in Fig.\,\ref{fig:WL1}\,{\bfE} stem from the gap region $\rg$ where we expect large values of $\vd$ to lead to full 0D confinement, we performed a spatial mapping of the amplitude of $X_\mathrm{0D}^{(1)}$ with energy $E$ of about $1637.5$\,meV at $\vd = 18$\,V. Fig.\,\ref{fig:WL1}\,{\bfF} shows that $X_\mathrm{0D}^{(1)}$ only appears when the optical spot with area $A_\mathrm{opt}$ overlaps with $\mathrm{R_G}$, thereby ruling out a contribution from extended 1D excitons along the finger gates. Further supporting evidence for confinement in $\mathrm{R_G}$ is provided by an additional measurement of $\Delta R/R_0$ conducted at the same location as a function of $\vs$, while keeping $\vlg = \vrg$ \cite{som}. Such a gate scan did not exhibit redshifting resonances when $\vtg \simeq 0\,$V, confirming our conclusion that the requisite electric fields are generated exclusively in $\mathrm{R_G}$, provided $\vd \ge 12$\,V. 


\begin{figure*}[ht!]
	\includegraphics[width=18.3cm]{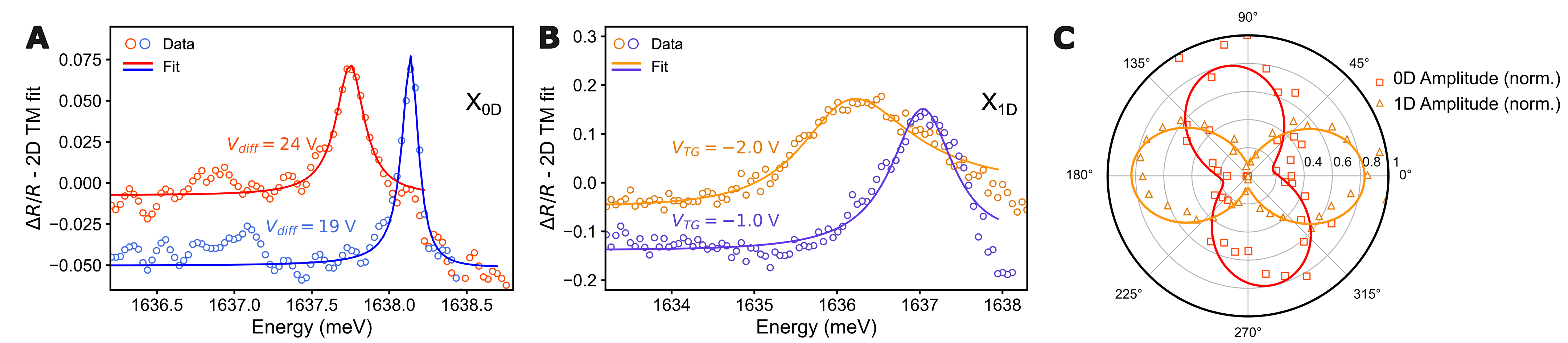}
	\caption{\textbf{Properties of quantum confined excitons in the gap region $\mathbf{R_G}$.}
    (\bfA) Line cuts of reflection spectra exhibiting only $\xdot$ states for fixed $\vtg = -0.4\,$V, $\vs=0\,$V, $V_\mathrm{diff,1} = 19\,$V (blue circles) and $V_\mathrm{diff,2} = 24\,$V (red circles) are obtained by subtracting the transfer matrix (TM) spectral profile of the 2D exciton background resonance $\xfree$. TM fits (solid lines) to these $\xdot$ resonances yields radiative decay rates $\hbar \gamma_\mathrm{rad,0D}$ of $(2.6\pm0.4)\,\mu$eV and $(3.0\pm0.3)\, \mu$eV, respectively. (\bfB) Line cuts of reflection spectra exhibiting only 1D quantum confined excitons for fixed $\vrg=12\,$V, $V_\mathrm{TG,1} = -1\,$V (purple circles) and $V_\mathrm{TG,2} = -2\,$V (gold circles) obtained in the same way. The narrow spatial region over the finger gate forms an extended electron-doped wire, while the remaining monolayer is hole-doped. A TM fit of these $\xwire$ resonances (solid lines) results in $\hbar \gamma_\mathrm{rad} = (40\pm10)\,\mu$eV.  (\bfC) By extracting the amplitude of $\xdot$ (red squares) and $\xwire$ (orange triangles) resonances in polarization-resolved $\Delta R/R_0$ measurements, we find $\xwire$ states to be polarized along the finger gates ($x$-axis) and $\xdot$ states perpendicular to them ($y$-axis). Solid lines indicate fits to the amplitude evolution, from which we extract a degree of linear polarization of $81\%$ for the $\xwire$ and $63\%$ for the $\xdot$ resonance.
	} 
	\label{fig:WL2}
\end{figure*}

The oscillator strength $f_\mathrm{0D}$ of an excitonic QD, where the confined COM wave function extends over an area $A_\mathrm{QD}  \ll  A_\mathrm{opt}$, is reduced with respect to that of the 2D exciton ($f_\mathrm{2D}$). An estimation of $f_\mathrm{0D}$ can be obtained by considering the Fourier transform of the normalized 0D exciton COM wave function $\psi_\mathrm{0D}(k)$ and calculating the probability $p$ of finding an exciton in this state within the free-space light cone, with $p = \int_\mathrm{light~cone}  |\psi_\mathrm{0D}(k)|^2 d^2k$. Our electrostatic simulations yield $p\simeq1/150$ for $\vd = 20\,$V and $\vtg = 0\,$V. To determine $f_\mathrm{0D}$ from our data, we first fit the 2D exciton spectral profile using a transfer matrix (TM) method, then subtract out its contribution from the reflectance spectrum and finally perform a second TM fit to the 0D resonance. Figure\,\ref{fig:WL2}\,{\bfA} shows the corresponding fits to $X_\mathrm{0D}^{(1)}$ resonance at different $V_\mathrm{diff}$ (see \cite{som} for details).  We find an average value of  $\hbar\gamma_\mathrm{rad,0D}\approx3.0\,\mu$eV in the  range of $V_\mathrm{diff}$ between $18\,$V and $24\,$V where $X_\mathrm{0D}^{(1)}$ is particularly well-resolved. To obtain the actual free space radiative decay rate, we need to take into account the fact that $A_\mathrm{opt}$ is larger than the optical scattering cross section of the 0D emitter~\cite{Karrai2003} $3\lambda_\mathrm{X}^2/2\pi$ at the emission wavelength $\lambda_\mathrm{X}=756\,$nm. Defining $\alpha_\mathrm{s}=3\lambda_\mathrm{X}^2/2\pi A_\mathrm{opt}$, we obtain the free-space radiative decay rate as $\gamma_\mathrm{rad,0D}/\alpha_\mathrm{scatter}$. Taking into account $\hbar\gamma_\mathrm{rad,2D}\approx1.2\,$meV obtained for the 2D exciton, this indicates a ratio $r_\mathrm{0D} = f_\mathrm{0D}/f_\mathrm{2D} = \gamma_\mathrm{rad,0D} / \alpha_\mathrm{s}\gamma_\mathrm{rad,2D}\simeq 1/170$, which is in good agreement with the probability $p$ we estimate using our electrostatic simulations. We remark that a slight energy difference of $X_\mathrm{0D}^{(1)}$ between Fig.\,\ref{fig:WL2}\,{\bfA} and Fig.\,\ref{fig:WL1}\,{\bfE} stems from a continuous slow shift of the resonance energies that occurs in our device upon its prolonged illumination, which can only be undone by thermal cycling.

To distinguish the observed resonances from 1D excitons that appear due to in-plane electric field gradients along a single direction, we also extracted radiative decay rates of 1D excitons from Fig.\,\ref{fig:WL1}\,{\bfB}. Employing a similar TM fitting procedure (Fig.\,\ref{fig:WL2}\,{\bfB}), we obtain an average $\hbar\gamma_\mathrm{rad,1D}\approx40\ \mu$eV for $-2\ \mathrm{V}<V_\mathrm{TG}<-0.5\ \mathrm{V}$, which reveals that the oscillator strength is only a factor of $r_\mathrm{1D} = f_\mathrm{1D}/f_\mathrm{2D} = \gamma_\mathrm{rad,1D}/ \sqrt{\alpha_\mathrm{s}}\gamma_\mathrm{rad,2D} \simeq 1/20$ smaller than that of the 2D exciton. The larger oscillator strength of the 1D exciton as compared to $X_\mathrm{0D}^{(1)}$ is consistent with the large spatial extent of the former along the finger gates.

Reflection measurements conducted on $R_\mathrm{G}$ further indicate that reducing $\vtg$ below $-1\,$V introduces hole doping in the gap region. We anticipate that such conditions may transform the confinement potential from being isotropic to being elongated along the $y$-direction \cite{som}. Due to this spatial anisotropy, we expect both the QD and 1D exciton resonances to be linearly polarized.
To confirm this, we carried out polarization-resolved reflection measurements where we excited the gap region using circularly polarized white light and measured the degree of linear polarization of the reflected light (Fig.\,\ref{fig:WL2}\,{\bfC}). We find that both $X_\mathrm{0D}^{(1)}$ and $X_\mathrm{0D}^{(2)}$  were polarized along the $y$-axis that is orthogonal to the axis of the finger gates, indicating that the confinement is weaker along $y$-direction. In contrast, we find the 1D excitons to be polarized along the $x$-direction, indicating that the strong confinement of 1D excitons is  along the $y$-direction, i.e. perpendicular to the axis of the finger gates, in agreement with earlier reports \cite{Thureja2022,Wang2001,Akiyama1996,Lefebvre2004,Bai2020,Wang2021,Heithoff2023}.

Surprisingly however, we only observe $y$-polarized 0D resonances, even for $\vd$ and $\vtg$ values that should yield relatively symmetric confinement potentials. To obtain further insight, we studied the degree of polarization as a function of external magnetic field $B_z$. Remarkably, we found that already for $B_z = 2.5\,$T the 0D exciton line was nearly circularly polarized \cite{som}. 
Assuming an exciton $g$-factor of $\sim4$ \cite{Srivastava2015} we extract an $x$-$y$ polarization splitting below $0.5\,$meV, consistent with nearly symmetric confinement.The absence of the second higher energy resonance with opposite polarization in $\Delta R/R_0$ for all $B_z$ field strengths remains an open question. Nonetheless, such behavior is consistent with findings pertaining to 1D localized excitons in monolayer WSe$_2$ \cite{Wang2021}.


The resonances $X_\mathrm{0D}^{(1)}$ and $X_\mathrm{0D}^{(2)}$ illustrated in  Fig.\,\ref{fig:WL1}\,{\bfE} and Fig.\,\ref{fig:WL2}\,{\bfA} exhibit three distinct characteristics: (i) a pronounced redshift with $\vd$, (ii)  spatial confinement within the gap region $\mathrm{R_G}$, constrained by the resolution of our imaging system, and (iii) strong linear polarization together with a linewidth $\delta \nu_\mathrm{QD,1} \lesssim 0.2\,$meV that is an order of magnitude narrower than that of the 2D exciton. These features strongly support~\cite{Hu2023} -- but do not prove -- their identification as stemming from full three-dimensional confinement of excitons. 

In prior experiments, 0D nature of tightly confined excitons in chemically synthesized nanocrystals or self-assembled QDs were established either through blinking/spectral wandering \cite{Nirmal1996,Empedocles1996,Robinson2000, Klein2019} or photon antibunching \cite{Michler2000a,Michler2000}. 
Spectral wandering stems from charge fluctuations that are ubiquitous in the solid-state  that generically result in time-dependent changes in the local potential experienced by excitons. While these fluctuations lead to inhomogeneous line broadening for 1D or 2D excitons, their effect on 0D excitons could be much more pronounced. Particularly, a single defect situated within few tens of nanometers from the 0D exciton will cause spectral jumps in the resonance fluorescence (RF) spectrum that can well exceed the exciton linewidth.

\begin{figure*}[ht!]
	\includegraphics[width=12.1cm]{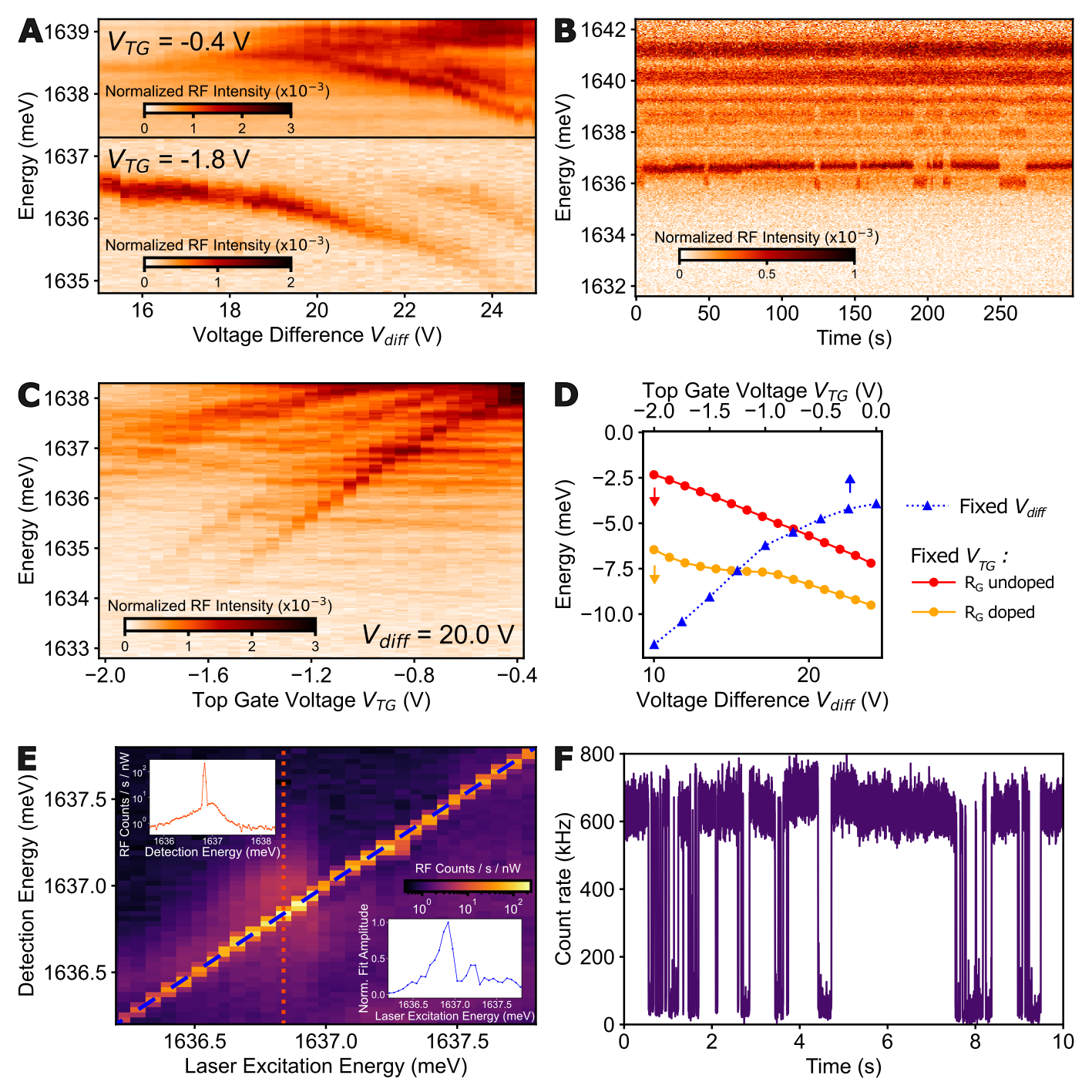}
	\caption{\textbf{Signatures of 0D quantum confined excitons in resonance fluorescence (RF).}
    (\bfA) Evolution of the normalized RF spectrum under white-light excitation for increasing $\vd$. At $\vtg=-0.4\,$V (upper panel) the extended MoSe$_2$ region is charge-neutral and becomes weakly hole-doped at $\vtg=-1.8\,$V (lower panel), which causes an overall redshift of the $\xdot$ states by approximately $2\,$meV. (\bfB) A time trace of the white-light RF at $\vtg=-1.8\,$V and $\vd=19\,$V displays spectral jumps in the energy of the $X_\mathrm{0D}^{(1)}$ state of around $0.6\,$meV. (\bfC) For fixed $\vd=20\,$V and decreasing $\vtg$, a pronounced redshift of the $\xdot$ states can be observed in the normalized white-light RF spectra. This behavior is in agreement with the evolution of the $X_\mathrm{0D}^{(1)}$ resonance energy, as determined by electrostatic simulations (blue triangles), shown in (\bfD). Similarly, our simulations also capture the global energy shift in the $\vd$-dependence of the $X_\mathrm{0D}^{(1)}$ resonance, when $\vtg$ is fixed such that $\mathrm{R_G}$ is charge-neutral (red circles) or lightly hole-doped (orange circles). (\bfE) To study the coherence properties of the QD, RF spectra are acquired using single-frequency laser excitation and sweeping its energy across the $\xdot$ resonances. As the laser becomes resonant with the $X_\mathrm{0D}^{(1)}$ energy at $\vtg=-1.8\,$V and $\vd=18\,$V, a clear increase in the detected RF signal is observable, as displayed in the lower right inset. The signal along the blue dashed line is dominated by the coherent Rayleigh scattering contribution to RF. In addition, the detected RF spectrum (along the orange dashed line) also features a broad incoherent scattering contribution (upper left inset). (\bfF) A time trace of the resonant-laser RF detected using a SSPD at $\vtg=-1.95\,$V and $\vd=21\,$V displays near complete blinking, possibly due to spectral jumps induced by charging and/or discharging of a trap state in close proximity to the confined exciton.
	} 
	\label{fig:RF}
\end{figure*}

To obtain a confirmation of the QD nature of $X_\mathrm{0D}^{(1)}$ and $X_\mathrm{0D}^{(2)}$ through observation of spectral wandering and/or blinking, we carry out RF measurements by exciting the device resonantly with light polarized along $(\hat{x} + \hat{y})/\sqrt{2}$ and detect the reflected signal polarized along $(\hat{x} - \hat{y})/\sqrt{2}$. This detection scheme allows us to eliminate the background from the 2D exciton, which is not possible in a PL experiment. In this manner, this configuration allows us to assess the degree of blinking in RF as a function of observation time. 
Fig.\,\ref{fig:RF}\,{\bfA} shows the white-light RF spectrum as a function of $\vd$  when the MoSe$_2$ layer (away from the finger gates) is either neutral ($\vtg= -0.4\,$V) or is lightly hole-doped ($\vtg= -1.8\,$V). Fig.\,\ref{fig:RF}\,{\bfB} shows a long time scan of RF we obtained for $X_\mathrm{0D}^{(1)}$ at $\vtg = -1.8\,$V and $\vd = 19\,$V. We observe that the $X_\mathrm{0D}^{(1)}$ resonance energy jumps between two values that differ by $0.6\,$meV on timescales of around $10\,$s. Repeating the measurement for different values of $\vtg$ and $\vd$, we find that spectral jumps are generic to this QD, even though the magnitude of the jumps varied substantially between different measurements.

We observe in Fig.\,\ref{fig:RF}\,{\bfA} that the 0D resonance redshifts by approximately $2\,$meV as finite background hole-doping is introduced. To confirm this finding we fixed $\vd = 20\,$V and measured the RF spectrum as a function of $\vtg$ (Fig.\,\ref{fig:RF}\,{\bfC}). The clear red shift of the 0D resonances with increasing $|\vtg|$ suggests that increasing background hole doping modifies the confinement potential in a way to increase the maximum electric field. Fig.\,\ref{fig:RF}\,{\bfD} shows the results of electrostatic simulations that are in excellent agreement with these observations. In particular, we find that in the presence of hole doping, the width of the neutral region along the $x$-direction is reduced while that along the $y$-direction is increased, rendering the confining potential more asymmetric \cite{som}. 

The coherence properties of the QD resonances could be studied by exciting the system with a single-frequency laser and monitoring the spectrum of the generated RF. To this end, we carried out complementary RF measurements where we scanned a single frequency laser across the $X_\mathrm{0D}^{(1)}$ resonance and monitored the full RF spectrum. Fig.\,\ref{fig:RF}\,{\bfE} shows the strength of RF as a function of the excitation and detection frequency after suppression of the incident laser field by cross-polarized detection. When the laser is resonant with the $X_\mathrm{0D}^{(1)}$ resonance, we observe a clear increase in the total detected RF signal (lower right inset). The RF spectrum obtained when the laser is on resonance in turn shows that in addition to coherent Rayleigh scattering, there is a sizeable incoherent light scattering contribution to RF (upper left inset). However, this incoherent contribution to RF does not have the same lineshape as the one observed in white-light RF, suggesting that it does not originate from a Markovian pure dephasing process. Understanding the nature of incoherent emission could provide insight into the nature of interactions between confined excitons and their electromagnetic environment.

Upon measuring the total RF counts using a superconducting single-photon detector (SSPD) as a function of time, we find near complete blinking in the detected RF signal (Fig.\,\ref{fig:RF}\,{\bfF}) -- presumably originating from spectral jumps taking place on timescales of few seconds. We tentatively attribute the abrupt changes in resonance energy to charging and/or discharging of a single trap in the very close vicinity of the confined exciton. Observation of telegraph-noise-like spectral jumps in white-light RF and blinking in resonant-laser RF together constitute a very strong indication of full quantum confinement of the excitons. 

In addition, the QD exciton experiences an electromagnetic environment from distant charge traps which introduce electric field fluctuations on timescales that are slower than the radiative lifetime but faster than the integration time.  We argue that the measured QD linewidths that significantly exceed the lifetime broadening are likely to originate from these fluctuations~\cite{Kuhlmann2013}. This assumption is further corroborated by the RF spectrum, shown in Fig.\,\ref{fig:RF}\,{\bfE}, where the coherent Rayleigh scattering is dominant; we note that quasi-static fluctuations should not lead to incoherent light emission. The finite incoherent emission background in Fig.\,\ref{fig:RF}\,{\bfE} in turn, is likely to originate either from pure dephasing due to fast electric field fluctuations or from phonon scattering~\cite{loudon2000quantum}.


While temporal spectral fluctuations would also cause  broadening of optical resonances of spatially extended (1D or 2D) emitters, we expect static spatial inhomogeneities arising from variations in local strain in this case to provide the dominant contribution to line broadening. A generic feature of 2D excitons for example is the variation in resonance energy and linewidth as the measurement position is changed by a few $\mu$m. In stark contrast, we observe a position-independent energy and linewidth (behold some small shifts due to light sensitivity, see \cite{som}) of the $X_\mathrm{0D}^{(1)}$ resonance when moving the excitation and detection spot as shown in Fig.\,\ref{fig:WL1}\,{\bfF}.



Photon antibunching in RF is considered an unequivocal signature of anharmonic quantum emitters in general \cite{Kimble1977}, and semiconductor QDs in particular \cite{Muller2007}. A QD for bosonic excitons that we study would exhibit an anharmonic spectrum leading to quantum correlations between generated photons if and only if exciton--exciton interaction strength $g_\mathrm{X-X}$ is large compared to the linewidth of the 0D quantum confined exciton mode. In contrast to our expectations, the photon correlation measurements we performed did not reveal a deviation from an uncorrelated Poissonian stream of RF photons. This observation suggests that  $g_\mathrm{X-X} \le 0.01\,\mu$eV$\mu$m$^2$, which is about two orders of magnitude smaller than the theoretical calculations \cite{Ciuti1998,Shahnazaryan2017}. We speculate that the fluctuations in the electric environment of the QD that lead to spectral jumps may be inhibiting the observation of photon antibunching. It is therefore imperative that improvements in sample design and manufacturing have to be achieved to obtain temporally stable and narrow spectral lines, which will be important to demonstrate emission of single indistinguishable photons. 

Our approach to achieve full quantum confinement can easily be adopted to other material systems like dipolar, layer-hybridized excitons in bilayer semiconductors \cite{Shimazaki2020,Sung2020,Tang2020}, and it can be integrated into electro-optical devices \cite{Wang2019,Uppu2021}. Realization of an array of identical quantum emitters by extending our findings should pave the way for a new class of applications, in particular in photonic quantum information processing. Furthermore, the bosonic QD hosting dipolar excitons that we demonstrate constitutes a new kind of quantum emitter with unique properties that deserve further investigation as a new playground for solid-state quantum optics.

\textbf{Acknowledgements}
This work was supported by the Swiss National Science Foundation (SNSF) under Grant Number 200020$\_$207520. The Authors acknowledge useful discussions with H. Adlong, X. Marie, C. Robert and A. Tugen. \textbf{Author contributions:} D.T., E.Y., and T.S. carried out the measurements and performed the data analysis. D.T. designed and fabricated the sample. D.T. and E.Y. carried out the numerical simulations. M.K. advised on the experiment and data analysis. D.T., E.Y., T.S., M.K., and A.I. wrote the manuscript. M.K., D.J.N., and A.I. supervised the project. \textbf{Competing interests:} The authors declare no competing interests. 

\newpage
\bibliography{References/Refs_v1.bib}

\clearpage

\section{Supplementary Information}
\renewcommand{\figurename}{SI Fig.}
\setcounter{figure}{0}
\renewcommand{\thefigure}{S\arabic{figure}}


\subsection{Sample fabrication}

The fabrication of split finger gate electrodes on SiO$_2$/Si substrates is achieved through electron beam lithography, followed by the electron beam evaporation of a Ti/Au layer ($3\,$nm/$10\,$nm). VdW heterostructures are fabricated using MoSe$_2$, h-BN and few layer graphene (FLG) flakes, which are obtained by mechanical exfoliation of bulk crystals (MoSe$_2$ from HQ Graphene, h-BN from 2D Semiconductors and FLG from NGS Naturgraphit) onto pristine $285\,$nm SiO$_2$/Si substrates using Scotch tape and Ultron wafer dicing tape. The heterostructure is assembled using a dry transfer technique \cite{Pizzocchero2016}, for which a hemispherical polydimethylsiloxane (PDMS) stamp is attached onto a glass slide. This stamp is thinly coated with polycarbonate (PC), which allows the sequential pickup of the flakes. The full vdW heterostructure consists of a monolayer MoSe$_2$ flake encapsulated by h-BN spacers of approximately $20-25$\,nm thickness. An extended top gate (TG) that covers the entire TMD region is made from a FLG flake. Additionally, the device features two contacts, also made with FLG, to the MoSe$_2$ monolayer, which allow for independent control over electron and hole injection in the device.
All stacking operations are conducted in an inert argon environment within a glovebox, at a temperature of $120$\,$^{\circ}$C. The completed stack is deposited onto the pre-patterned substrate containing the bottom gate electrodes by raising the temperature to $150$\,$^{\circ}$C. This increase in temperature enables the PC to detach from the PDMS and adhere to the substrate. Further heating to $170$\,$^{\circ}$C induces tearing of the PC layer at the edges. To dissolve the PC, the substrate is immersed in chloroform. Finally, the device is completed by forming metal electrodes for the contacts and gates with Ti/Au ($5\,$nm/$85\,$nm). 

\subsection{Experimental setup}

To conduct our optical experiments we make use of a confocal microscope setup. The sample is mounted on piezoelectric nanopositioners within a stainless steel tube, which is submerged in a liquid helium bath cryostat. This arrangement maintains the sample at approximately $4.2$\,K, facilitated by the introduction of $20$\,mbar helium exchange gas into the tube. A glass window on top of the tube provides free-space optical access to the sample. White-light reflectance measurements are performed by employing a broadband light-emitting diode (LED) centered at $760$\,nm as an excitation source. For resonant measurements a single-mode wavelength-tunable Ti:sapphire laser is utilized. The excitation light is focused onto the sample via a high-numerical-aperture lens ($0.68$), creating a diffraction-limited spot. The reflected or emitted light from the sample is collected using the same lens, separated from the incident light by a beam splitter, guided into a single-mode fibre and imaged on a spectrometer equipped with a liquid-nitrogen-cooled charge-coupled device. An excitation power of a few tens of nW is maintained during both white-light and broadband resonance fluorescence (RF) measurements. In the latter case, a set of polarization optics, including a linear polarizer and a quarter-wave plate, is placed in both excitation and detection paths. This allows us to ensure a sizable suppression ($\gtrsim10^5$) of reflected light that is co-polarized with respect to the linearly-polarized excitation beam. In order to control the orientation of this linear polarization during the experiments, we rotate a half-wave plate that is placed in a common path (i.e.\,before the beam splitter that separates exciting and reflected beams).

The same setup is also used for narrow-band RF measurements employing a single-frequency tunable laser as well as for single-photon-correlation experiments. In the latter case, the excitation power is raised to up to $\sim\mu$W to increase the photon count rate. Moreover, the RF signal is transmitted in a monomode fiber to a Hanbury Brown and Twiss setup (based on a fiber-beam-splitter) that splits it into two, almost equally-intense beams. The correlation between photons in each of these beams are measured with a pair of fast ($\sim8\,$ps), fiber-coupled superconducting single photon detectors (SSPDs) operated at a temperature $<10\,$K in a separate, closed-cycle cryostat.

\subsection{Electrostatic simulation of device geometry}

\begin{figure*}
    \includegraphics[width=12cm]{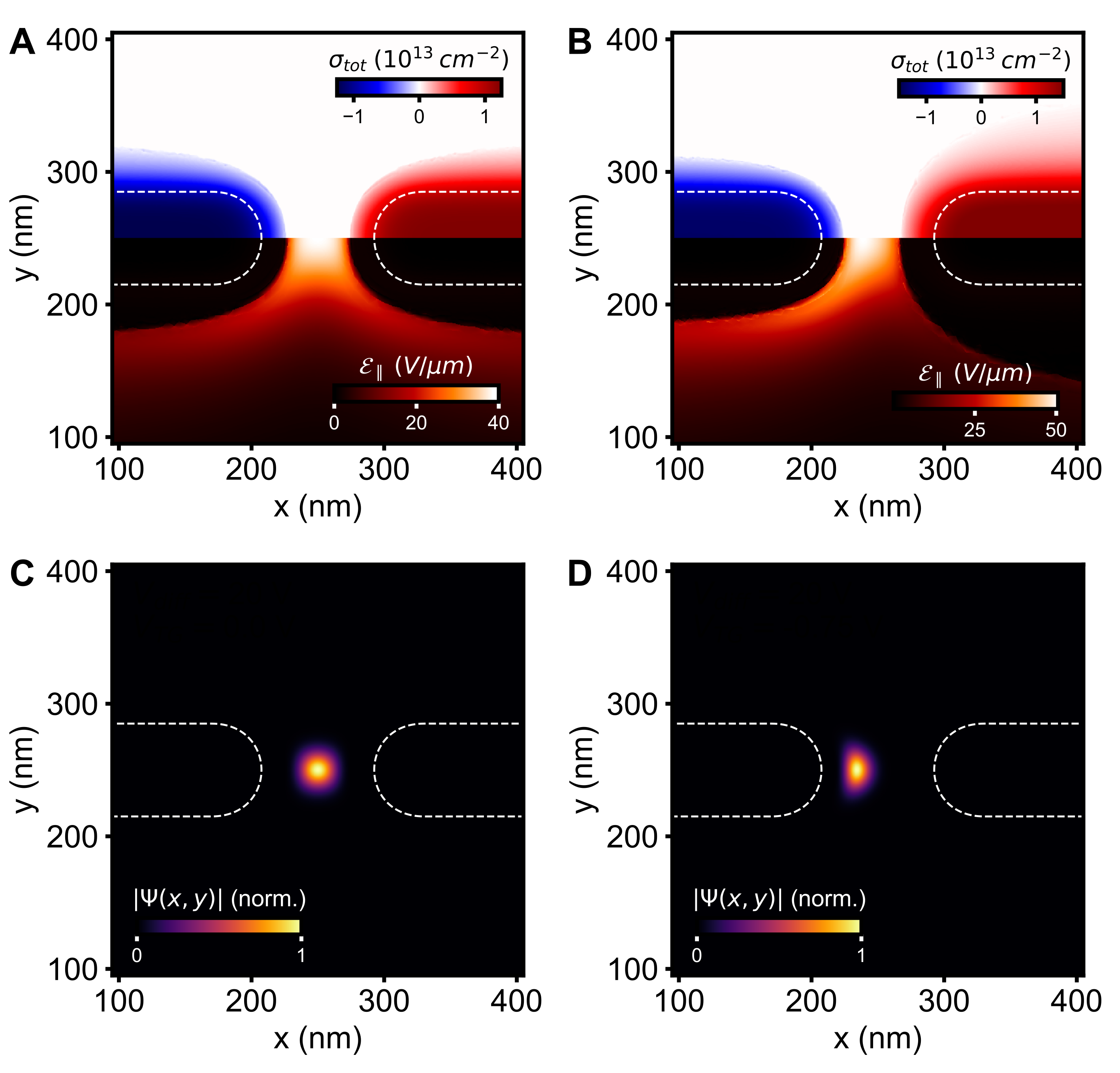}
    \caption{\textbf{0D exciton simulation results.} (\bfA) and (\bfB) show the electrostatics simulation results for the undoped and slightly background doped cases at fixed $\vd$. Upper part of both panels is the charge density and the lower part is the in-plane electric field magnitude in the monolayer. White dashed lines indicate the location of the finger gates. (\bfC) and (\bfD) show the resulting wavefunction of the confined exciton given the electrostatic environment in panels (\bfA) and (\bfB) respectively. The wavefunctions are normalize to their maximum value for clarity. The extent of the wavefunctions are $(\sigma_x, \sigma_y)$ = (9.1nm, 9.1nm) in (\bfC) and (6.4nm, 10nm) in (\bfD).
    }
    \label{fig:supp_sim}
\end{figure*}

The calculation of in-plane electric field and charge density distributions presented in the main text are performed according to the procedures described in Refs.\,\cite{Thureja2022,Thureja2023}. These finite-element calculations assume the following material parameters: monolayer MoSe$_2$ bandgap $E_\mathrm{g} = 1.85$\,eV, Fermi level offset relative to the valence band edge at zero potential $E_\mathrm{F} - E_\mathrm{V}(V=0)=0.99$\,eV \cite{Wilson2017}, electron effective mass $m_\mathrm{n}^* = 0.7\,m_\mathrm{e}$ \cite{Larentis2018}, hole effective mass $m_\mathrm{p}^* = 0.6\,m_\mathrm{e}$ \cite{Zhang2014,Goryca2019}, out-of-plane dielectric constant  $\varepsilon_{\perp} = 3.76$, and in-plane dielectric constant $\varepsilon_{\parallel} = 6.93$ for h-BN \cite{Laturia2018}. Having determined the electrostatic landscape in the device, i.e.\,the in-plane electric field $F_{\parallel}$ and the charge density $\sigma$, we calculate the excitonic confinement potential:
\begin{equation}
    V(x,y) = \underbrace{-\frac{1}{2}\alpha |F_{\parallel}(x,y)|^2}_\text{dc Stark shift} \,\,\,\, + \underbrace{\beta |\sigma(x,y)|}_\text{Interaction shift}
    \label{eqn:potential}
\end{equation}
where we assume an exciton polarizability $\alpha = 6.5\,\mathrm{eV\,nm}^2/\mathrm{V}^2$ \cite{Cavalcante2018} and an effective exciton--charge coupling strength $\beta \simeq 0.7\,\mu\mathrm{eV}\mu\mathrm{m^2}$ \cite{Thureja2022}. 

Naturally, the electrostatics depend on the voltages applied on the finger gates and the global top gate. 
Larger $\vd$ values result in stronger in-plane fields created in $\rg$ while introducing background electron (hole) doping narrows the charge neutrality region and shifts it towards the negatively (positively) biased finger gate.
As can be seen in Figure \ref{fig:supp_sim}, we can solve the eigenvalue problem and extract the extent of the COM wavefunction in both cases to be firmly in the quantum confined regime: $\sigma_{x},\, \sigma{y}\leq 10$nm $\ll h/\sqrt{2m_X k_B T}$.

If too much doping is introduced, screening in the gap region due to background charges becomes important and the neutrality region wraps around the finger gate that is biased opposite to global doping.
In this case, the confined exciton ground state is closer to a 1D wire with tight confinement and the associated redshift is much larger than that of a 0D exciton with symmetric confinement.


\subsection{Device doping characteristics}

In this section, we discuss the charging behaviour of our device in greater detail and determine a gate voltage configuration that ensures the observation of 0D quantum confined excitons. To this end, we measure normalized differential reflectance $\Delta R/R_0$ on the region $\mathrm{R_L}$ of Device 1 and perform a full dual-gate scan, where the gate voltages $\vtg$ and $\vlg$ are varied across their entire accessible range. Then, by mapping the differential reflectance $\Delta R/R_0$ at fixed energy $E = E_\mathrm{X,2D} + \Gamma_\mathrm{2D}/2 = 1642.5\,$meV the charging behaviour of the device can be characterized more rigorously. Here, $\Gamma_\mathrm{2D}$ is the 2D exciton linewidth. Since the optical spot is diffraction-limited, the measured optical response will contain information about the doping state of both the dual-gated and extended TMD regions. The emergence of a repulsive polaron resonance will thereby serve as an indicator for the onset of doping.

\begin{figure}[htb]
    \centering
	\includegraphics[width=7cm]{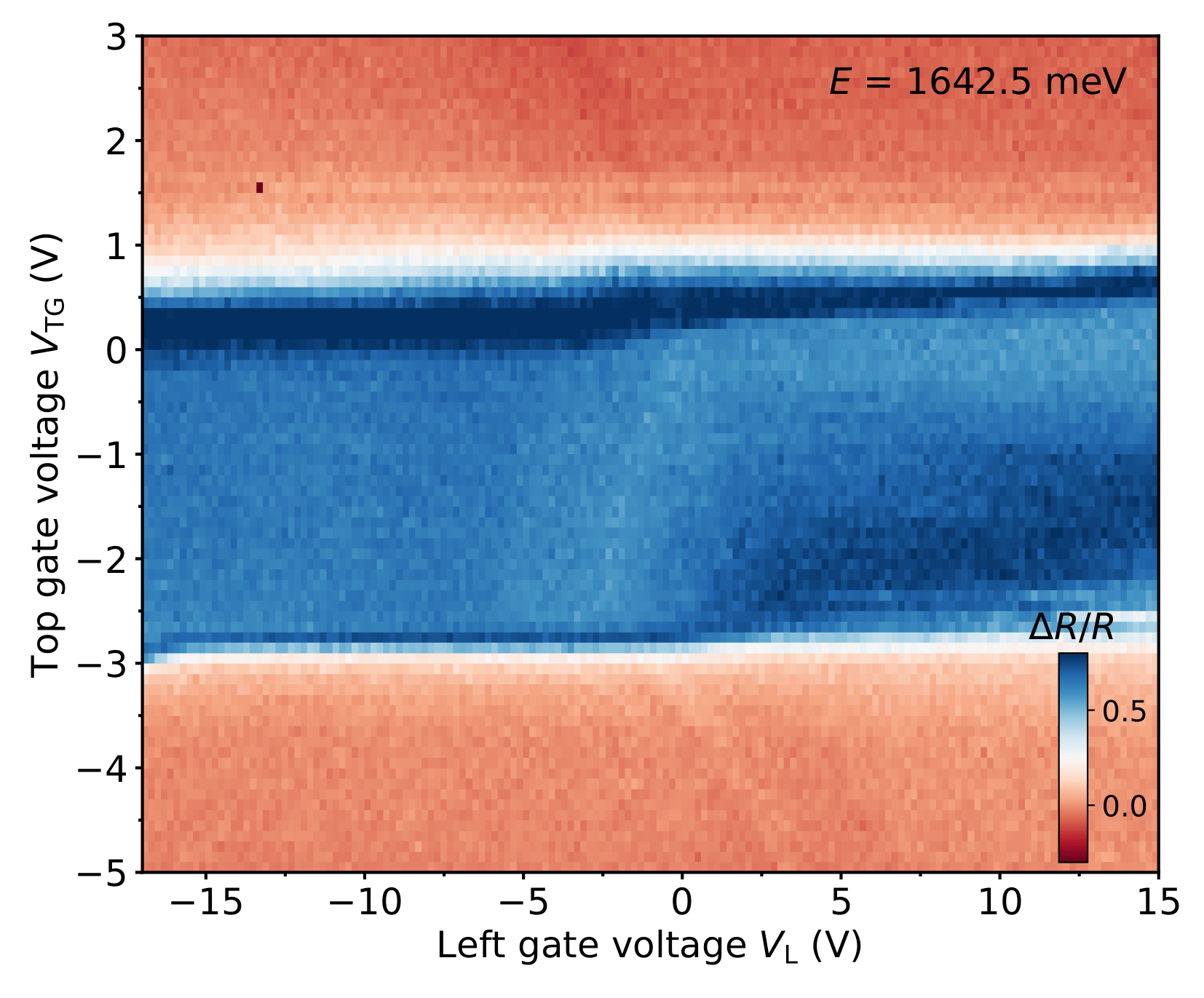}
	\caption{\textbf{Doping characteristics of Device 1.}
    We focus on charge-density-dependent energy shifts of the repulsive polaron to identify the doping configuration over the region $\mathrm{R_L}$ of Device 1. We measure normalized differential reflectance $\Delta R/R_0$, both as a function of $\vlg$ and $\vtg$, and plot the result at fixed energy, which is chosen slightly blueshifted to the bare 2D exciton i.e.\,at $E=1642.5\,$meV. The prominent horizontal lines demarcate the n-doped, neutral and p-doped regimes in the extended TMD region. The fainter set of resonances exhibiting a zigzag behaviour originate in the TMD region $\mathrm{R_L}$ and indicate onset of doping in this area. It can be clearly seen that for $\vtg$ ranging from approximately $-1\,$V to $0\,$V, $\vlg$ can be tuned across its entire range without greatly affecting charge neutrality in the extended TMD region.
    }  
	\label{fig:device1_doping}
\end{figure}

Fig.\,\ref{fig:device1_doping} depicts the result of such a dual-gate scan, where $\vtg$ and $\vlg$ are varied. Overall, the observed features match very well with the doping characteristics of the devices investigated in our prior study \cite{Thureja2022}. The prominent horizontal lines indicate onset of electron doping for $\vtg\gtrsim0.5\,$V and hole doping for $\vtg\lesssim-3\,$V in the extended TMD region. Concurrently, a fainter set of resonances exhibiting a zigzag behaviour is visible. These spectral lines point at the doping state in the region $\mathrm{R_L}$. To their right, for increasing $\vlg$, the back-gated area is electron-doped. To their left, for decreasing $\vlg$, it becomes hole-doped. Based on this charging map, the most desirable conditions for probing 0D states are for $-1\,\mathrm{V}\lesssim\vtg\lesssim0\,$V. Within this range the bottom-gated area can be both p-doped and n-doped, while least affecting charge neutrality in the extended TMD region, in good agreement with the desired doping configuration shown in Fig.\,\ref{fig:concept}\,{\bfC}. Such charging maps were also acquired for other split-gate electrodes (in Device 1 and Device 2) which showed similar doping behaviour.

\subsection{\texorpdfstring{$\vs$}{ }-dependent spectra}

\begin{figure*}
	\includegraphics[width=12cm]{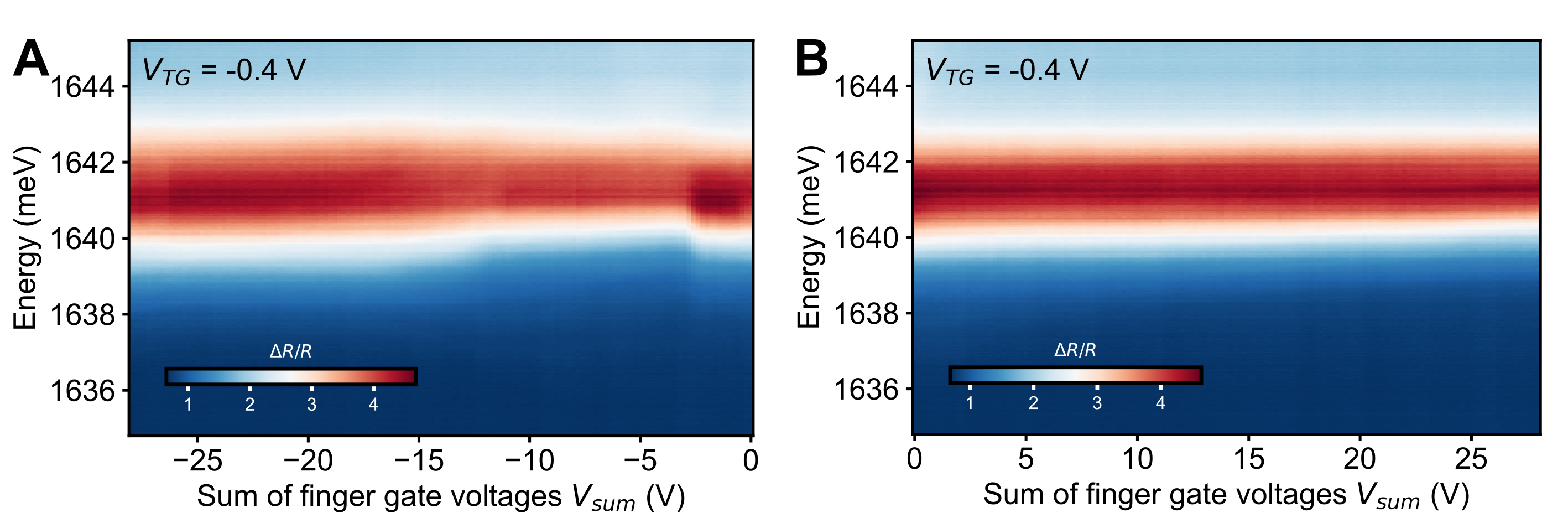}
	\caption{\textbf{Reflectance contrast spectra when finger gates are biased with the same polarity.} (\bfA) In the negative $\vs$ direction, we observe variations in the exciton resonance due to changing doping around the finger gates. There is no sign of a resonance with strong dependence on $\vs$. (\bfB) In positive direction there is little change in the reflectance for all values of $\vs$.} 
	\label{fig:supp_vsum}
\end{figure*}

In order to confirm that the resonances we attribute to 0D excitons arise due to asymmetry between the two finger gates and not due to that between one finger gate and the global top gate, we measured $\vs$-dependent reflectance contrast spectra on $\rg$. For these measurements, we biased the two gates symmetrically ($\vd = 0$). 
In this case, stray in-plane electric fields on the TMD should be small and thus no sign of our 0D quantum confined states are expected. 
In Figure \ref{fig:supp_vsum} we report our findings. 
As anticipated, there are no resonances that redshift with the applied voltage. 

\subsection{Normalization of reflectance spectra}

We normalize the reflectance spectra we present always with respect to a reference spectrum $R_0$ that is taken on the same spot, but with a large top-gate voltage ($> 7$ V) that highly dopes the monolayer TMD. 
The differential reflectance is then given by:
\begin{equation}
    \frac{\Delta R}{R_0} = \frac{R - R_0}{R_0}
\end{equation}
where $R$ is the bare reflectance of interest.

For white-light resonance-fluorescence measurements, the detected spectrum $R_\mathrm{RF}$ is normalized with respect to the reference spectrum $R_0^\mathrm{unsupp.}$ acquired at high top-gate voltage in the absence of RF suppression to account for the spectral shape of the excitation source.
\begin{equation}
    \text{Normalized RF intensity} = R_\mathrm{RF}/R_0^\mathrm{unsupp.}
\end{equation}
\subsection{Differentiation and fitting of reflectance spectra}

Although $\xdot$ is strongly dependent on the applied bias  $\vd$ between the two finger gates, $\xfree$ is not. This asymmetry gives us a powerful tool to resolve the spectral signatures of $\xdot$ despite the fact that it lies on the shoulder of sizably broader and orders of magnitude stronger $\xfree$ resonance. More specifically, we differentiate the data with respect to $\vd$, which allows us to isolate 0D resonances by suppressing the $\vd$--independent background (as seen in Fig.\,\ref{fig:WL1}). Techniques we discuss here are analogous to well-established procedures in modulation spectroscopy.

We followed the following differentiation process to acquire these spectra:
\begin{equation}
    \left( \frac{\partial}{\partial \vd} \frac{\Delta R}{R_0} \right) (V_i) := \frac{R(V_{i+\delta}) - R(V_{i-\delta})}{(V_{i+\delta}) - (V_{i-\delta})} \cdot \frac{1}{R_0}
    \label{eq:differentiation}
\end{equation}

\begin{figure*}
	\includegraphics[width=18cm]{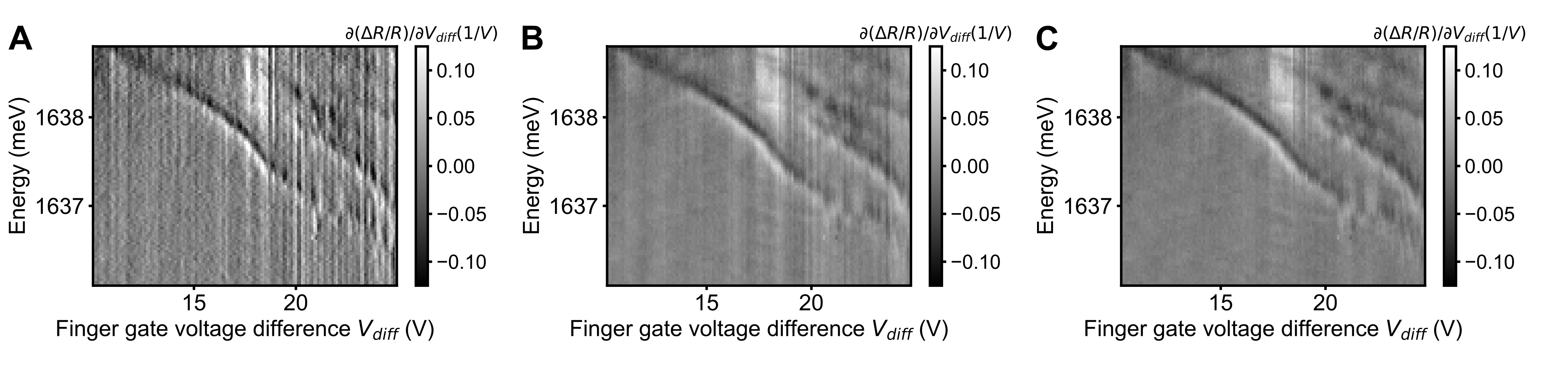}
	\caption{\textbf{Different ways to obtain differentiated reflectance contrast spectra.} (\bfA) Applying Eq.\,\ref{eq:differentiation} with $\delta = 1$ and without offset subtraction. (\bfB) Setting $\delta = 3$ amplifies the useful part of the differential signal while suppressing random noise. (\bfC) Subtracting the finite offset introduced by intensity fluctuations mostly eliminates vertical stripe-like artifacts.
    }
	\label{fig:supp_differentiation}
\end{figure*}

Here, $R_0$ is the reference spectrum acquired at large top gate voltage, $V_i$ denotes $\vd$ for data point $i$, while $R(V_i)$ is the corresponding bare reflectance spectrum. Setting $\delta$ larger than one enhances the signal-to-noise ratio by suppressing the high frequency noise (e.g. dark counts on the spectrometer). This can be seen in Figs\,\ref{fig:supp_differentiation}{\bfA},{\bfB} that show differentiated spectra from Fig.\,\ref{fig:supp_vsum} for $\delta=1$ and $\delta=3$.

Since the contribution of 0D resonances to $R(V_{i+\delta}) - R(V_{i-\delta})$ is very small, even small intensity fluctuations of the excitation source can lead to substantial change of $\partial(\Delta R/R_0)/\partial\vd$ that might be comparable in amplitude to the analyzed 0D spectral features. This effect manifests itself in irregular jumps of $\partial(\Delta R/R_0)/\partial\vd$ that we observe for some datasets (e.g., in Fig.\,\ref{fig:supp_differentiation}). In order to suppress them, we select a narrow spectral window that hosts no spectral resonances ($1620\,\mathrm{meV} < E < 1625\,\mathrm{meV}$). For each voltage $V_i$, we calculate the mean $\langle \partial(\Delta R/R_0)/\partial\vd\rangle$ in this spectral range and subtract it from the differentiated signal. As shown in Fig.\,\ref{fig:supp_differentiation}{\bfC}, this procedure can substantially reduce the background noise. We note, however, that it was used only for some supplementary figures and not for those in the main text.

\begin{figure}
	\includegraphics[width=6cm]{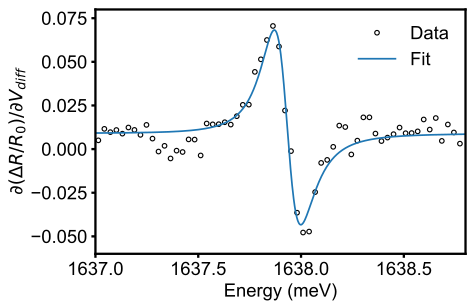}
	\caption{\textbf{Fitting differential reflectance spectra.}  Derivative of a Lorentzian with absorptive and dispersive contributions describe very well the differential reflectance data we obtained where the resonance energy is modulated by changing the applied voltage.
 }
 \label{fig:supp_diff_fit}
\end{figure}

To describe the resonances that appear in differentiated reflectance spectra, we fit their lineshapes with a derivative of a dispersive Lorentzian spectral profile $\partial\mathcal{L}_0 (E)/\partial E$ that is given by:
\begin{align}
    \mathcal{L}_0 (E) &= A \cdot \frac{\cos{\alpha} \cdot \Gamma/2 + \sin{\alpha} \cdot(E_0 - E)}{(E_0 - E)^2 + \Gamma^2/4}\\
    \frac{\partial}{\partial E}\mathcal{L}_0 (E) &= A\cdot \frac{-\cos{\alpha}\cdot\Gamma\,\Delta_E + \sin{\alpha}\cdot \left( - \Delta_E^2 + \Gamma^2/4 \right ) }{(\Delta_E^2 + \Gamma^2/4)^2}
    \label{eq:diff_lorentz}
\end{align}
where $E$ denotes the photon energy, $E_0$ is the resonance energy, $\alpha$ is the phase, $\Gamma$ is the linewidth, $A$ is the amplitude, while $\Delta_E = (E_0 - E)$. In Fig.\,\ref{fig:supp_diff_fit}, we show an example fit to the spectrum measured at $\vd=16.6\,$V in Fig.\,\ref{fig:WL1}\,{\bfE} inset.

\subsection{Determination of oscillator strength with transfer matrix method}

\begin{figure*}
	\includegraphics[width=16cm]{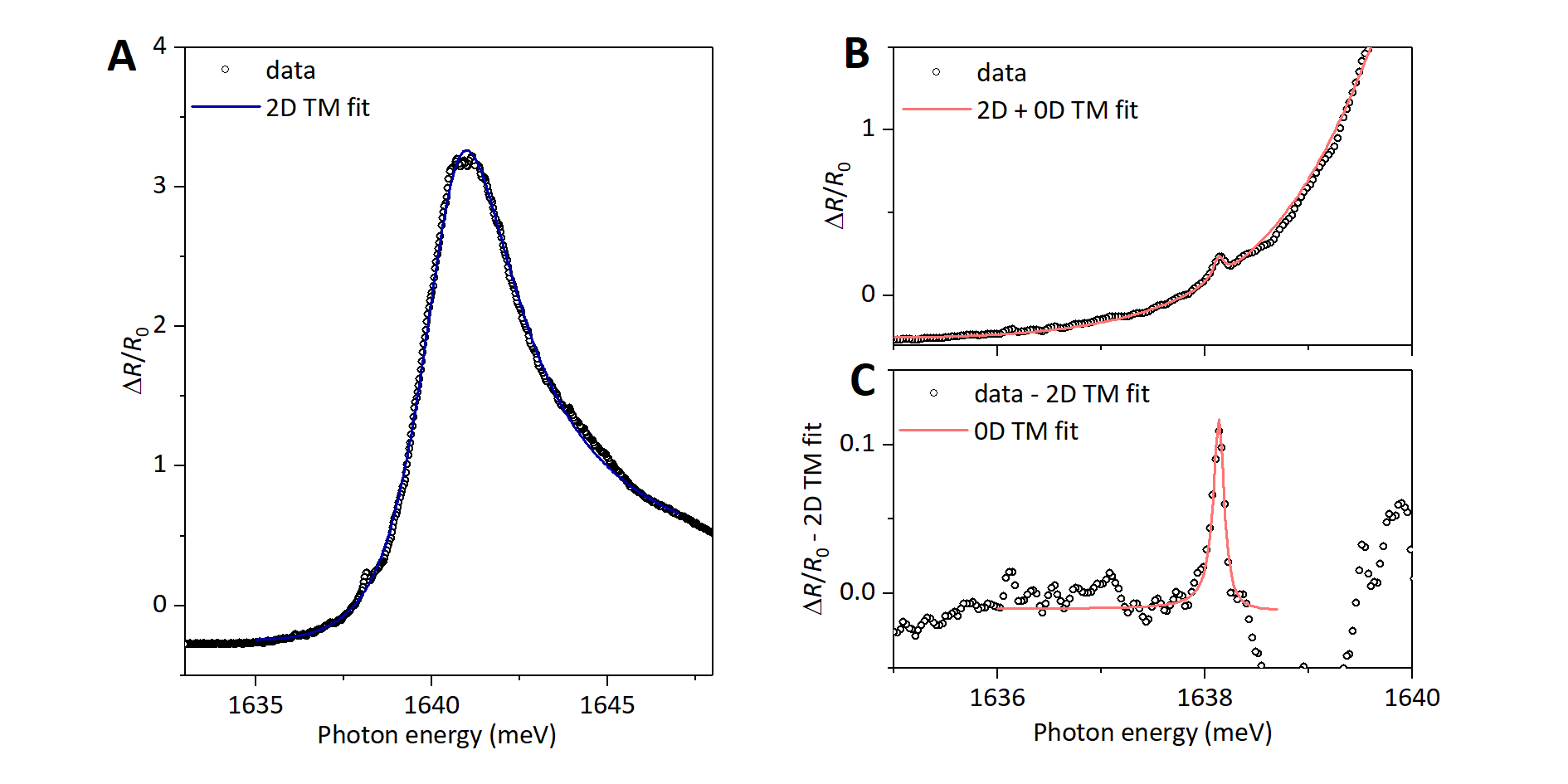}
	\caption{\textbf{Further data on the TM fitting procedure.} (\bfA) Reflectance contrast spectrum measured at $V_\mathrm{diff}=19\,$V in the gap region. The solid line represents the TM fit to the data corresponding to $\hbar\gamma_\mathrm{rad,2D}\approx1.2\,$meV and $\hbar\gamma_\mathrm{nrad,2D}\approx2.4\,$meV. (\bfB) The same data as in (\bfA), but this time fitted with a TM model (solid line) taking into account both 2D and 0D exciton resonances (the parameters for the former one are fixed). (\bfC) The same data and fit as in (\bfB), but with subtracted TM fit to the 2D exciton resonance.} 
	\label{fig:TM_supp}
\end{figure*}

In order to extract radiative decay rates of the exciton transitions, we fit our reflectance contrast spectra with a transfer matrix (TM) method~\cite{Back2018,Scuri2018,Smolenski2022}. To this end, we take the h-BN refractive index of 2.1~\cite{Lee2019} and assume that the thicknesses of top and bottom h-BN layers in our device are $22\,$nm and $24\,$nm, respectively (these values are consistent with atomic force microscopy results and allow us to get the best TM fits to our data).

Independent of the dimensionality of the analyzed excitonic resonance, the fitting procedure always begins with fitting the lineshape of a 2D exciton resonance in a given reflectance contrast spectrum. This is carried out under the assumption that the susceptibility of the MoSe$_2$ monolayer, which depends on the energy $E$, has the following form
\begin{equation}
    \chi_\mathrm{2D}(E)=-\frac{(\hbar^2c/E_\mathrm{X,2D})\gamma_\mathrm{rad,2D}}{(E-E_\mathrm{X,2D}+i\hbar\gamma_\mathrm{nrad,2D}/2)}
\end{equation}
Here, $\hbar$ is the reduced Planck constant, $c$ is the speed of light in vacuum, $E_\mathrm{X,2D}$ is the 2D exciton energy, $\gamma_\mathrm{nrad,2D}$ is the non-radiative exciton decay rate, while $\gamma_\mathrm{rad,2D}$ represents a free-space exciton decay rate, which is a rate at which the 2D exciton would radiatively decay if the MoSe$_2$ monolayer were placed in vacuum. We note that this does not coincide with the actual decay rate owing to the interference of light reflected off multiple interfaces in the device, which results in a Purcell effect~\cite{Fang2019}. Fig.\,\ref{fig:TM_supp}\,{\bfA} shows an example fit carried out in the gap region $\mathrm{R_G}$ at $V_\mathrm{diff}=19\,$V, yielding $\hbar\gamma_\mathrm{rad,2D}\approx1.2\,$meV and $\hbar\gamma_\mathrm{nrad,2D}\approx2.4\,$meV, which are typical for the 2D exciton in charge-neutrality in our device.

To fit the spectral profile of a $n\mathrm{D}$ ($n=0,1$) resonance, we fix the parameters of the 2D resonance and fit the same reflectance contrast spectrum again with the following susceptibility: $\chi\mathrm(E)=\chi_\mathrm{2D}(E)+\chi_{n\mathrm{D}}(E)$, where $\chi_\mathrm{2D}$ is the above-defined, fixed contribution from the 2D exciton resonance, while $\chi_{n\mathrm{D}}(E)=-(\hbar^2 c/E_{\mathrm{X,}n\mathrm{D}})\gamma_{\mathrm{rad,}n\mathrm{D}}/(E-E_{\mathrm{X,}n\mathrm{D}}+i\hbar\gamma_{\mathrm{nrad,}n\mathrm{D}}/2)$ represents the contribution of the analyzed $n$D resonance with an energy $E_{\mathrm{X,}n\mathrm{D}}$ and non-radiative decay rate of $\gamma_{\mathrm{nrad,}n\mathrm{D}}$. In analogy to the 2D exciton, $\gamma_{\mathrm{rad,}n\mathrm{D}}$ denotes the radiative decay rate of the $n$D exciton, which would correspond to its free-space value if the optical spot size $A_\mathrm{opt}$ would be equal the optical scattering cross-section $A^{n\mathrm{D}}_\mathrm{scatter}$ of the $n$D emitter. Since these two quantities are different in our experiments, the actual free-space radiative decay rate yields $\gamma_{\mathrm{rad,}n\mathrm{D}}A_\mathrm{opt}/A^{n\mathrm{D}}_\mathrm{scatter}$. In case of the 0D emitter, $A^{0\mathrm{D}}_\mathrm{scatter}=3\lambda_\mathrm{X}^2/2\pi$ (where $\lambda_\mathrm{X}$ is the 0D emission wavelength) and thus $A_\mathrm{opt}/A^{0\mathrm{D}}_\mathrm{scatter}=1/\alpha_\mathrm{s}=2\pi A_\mathrm{opt}/3\lambda_\mathrm{X}^2$. For the 1D emitter, the analyzed ratio is given by a square root of this coefficient, i.e., $A_\mathrm{opt}/A^{1\mathrm{D}}_\mathrm{scatter}\approx\sqrt{1/\alpha_\mathrm{s}}$.

It is important to note that the fit to 0D/1D resonances is carried out in a narrower energy range in the vicinity of the $n$D resonance, which is selected to avoid residuals of the 2D exciton fitting. Fig.\,\ref{fig:TM_supp}\,{\bfB} illustrates such a fit at $V_\mathrm{diff}=19\,$V that, after subtracting the fitted lineshape of the 2D exciton, corresponds to the data shown in Fig.\,\ref{fig:WL2}\,{\bfA} in the main text (also presented in Fig.\,\ref{fig:TM_supp}\,{\bfC}).

\subsection{Influence of magnetic field on the polarization of 0D resonance}

\begin{figure*}
	\includegraphics[width=13cm]{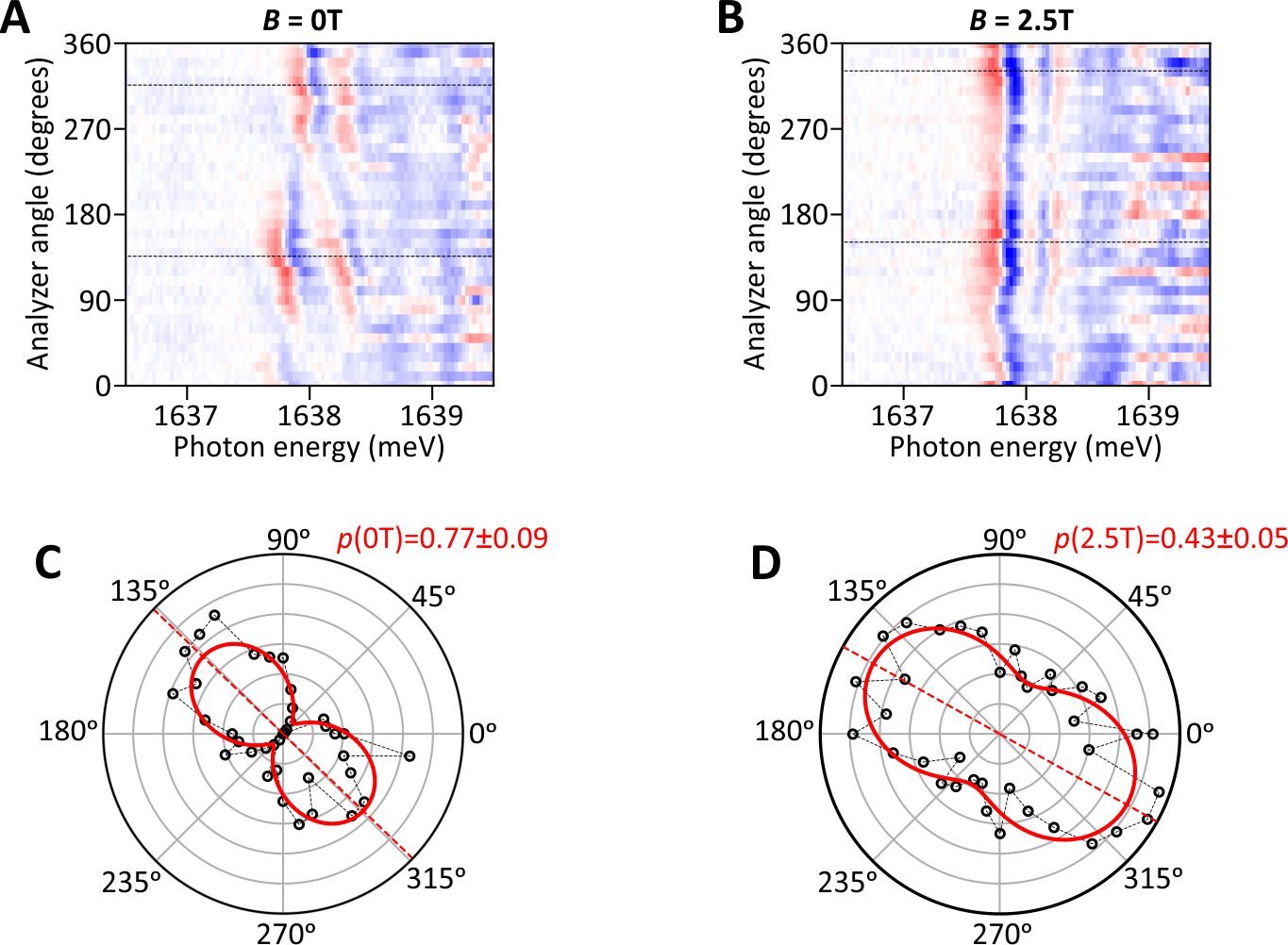}
	\caption{\textbf{Magnetic field evolution of 0D exciton polarization degree.} (\bfA-\bfB) Maps presenting differentiated reflectance contrast spectrum of a 0D exciton as a function of linear analyzer angle for two different magnetic fields: $0\,$T (\bfA) and $2.5\,$T (\bfB). The spectra were measured under circularly polarized excitation at $V_\mathrm{TG}=-0.4$~V and $V_\mathrm{diff}=22\,$V. Dashed lines indicate eigen-polarization orientation. (\bfC-\bfD) Polar plots showing the corresponding angular evolution of the amplitude of the lowest-energy 0D resonance extracted by fitting its spectral profile with differentiated dispersive Lorentzian curve. Solid lines represent the fits to the data with the following formula $I(\alpha)\propto[2p\cos^2(\alpha-\varphi_0)+(1-p)]$, where $\varphi_0$ represents the eigen-linear-polarization orientation (marked with a red dashed line), while $p$ is the linear polarization degree (indicated at each panel).}
	\label{fig:magnetic_pol_supp}
\end{figure*}

As noted in the main text, the 0D exciton lines we observe in our experiments are linearly polarized along the $y$-direction perpendicular to the orientation of the two finger gates. This finding is surprising given that the confining potential generated by these gates is expected to be rather symmetric, which should in turn result in the presence of two, orthogonally-polarized 0D lines split by the $x-y$ polarization splitting $\delta_1$, as it is typically the case for excitons in self-assembled quantum dots~\cite{Bayer2002}. 

To further understand the absence of $x$-polarized 0D transitions, we perform linear-polarization-resolved measurements of 0D excitons as a function of out-of-plane magnetic field. More specifically, we excite the sample with a circularly-polarized broadband white-light and acquire differentiated reflectance contrast spectra for different orientations of a linear analyzer. Figs.\,\ref{fig:magnetic_pol_supp}\,{\bfA},{\bfB} show such angle-dependent spectral evolutions measured both at zero $B$-field and at $B=2.5\,$T. Although the field that we apply generates only a small Zeeman shift for the 2D exciton of $g_\mathrm{X}\mu_\mathrm{B}B\approx0.5\,$meV (assuming $g_\mathrm{X}\approx4$~\cite{Srivastava2015}; $\mu_\mathrm{B}$ is a Bohr magneton), it sizably affects the polarization the 0D emitter: while the orientation of its polarization remains almost the same as at $B=0$, the polarization degree $p$ is clearly lower. To analyze this more quantitatively, we fit the spectral profile of our 0D resonance with a differentiated Lorentzian lineshape, which allows us to extract its angle-dependent intensity (Figs.\,\ref{fig:magnetic_pol_supp}\,{\bfC},{\bfD}). The polarization degree obtained on this basis indeed decreases by roughly 50\% upon application of $B=2.5\,$T, which would suggest that the polarization splitting $\delta_1\lesssim0.5\,$meV. While this finding is consistent with an almost symmetric confining potential, the lack of any spectral signatures of higher-energy $x$-polarized transitions becomes even more surprising in view of such a small $\delta_1$ splitting. Although a detailed understanding of this issue remains beyond the scope of this work, we note that similar polarization properties are sometimes displayed by 0D/1D emitters naturally emerging in 2D materials~\cite{Koperski2018,Wang2021}.

\subsection{Dependence of 0D exciton energy on optical excitation location}

Owing to its strong spatial confinement, the 0D exciton is especially sensitive to the local charge environment, as evidenced in Fig.\,\ref{fig:RF}\,{\bfB}. In particular, any changes in the electrostatic confinement landscape, due to e.g.\,modification of doping conditions, inherently result in variations of the $X_\mathrm{0D}$ resonance energy. Therefore, spectral signatures of the 0D exciton depend on the optical excitation setting. In particular, in our device $X_\mathrm{0D}$ resonance typically blueshifts with increased excitation power.

\begin{figure*}
	\includegraphics[width=18cm]{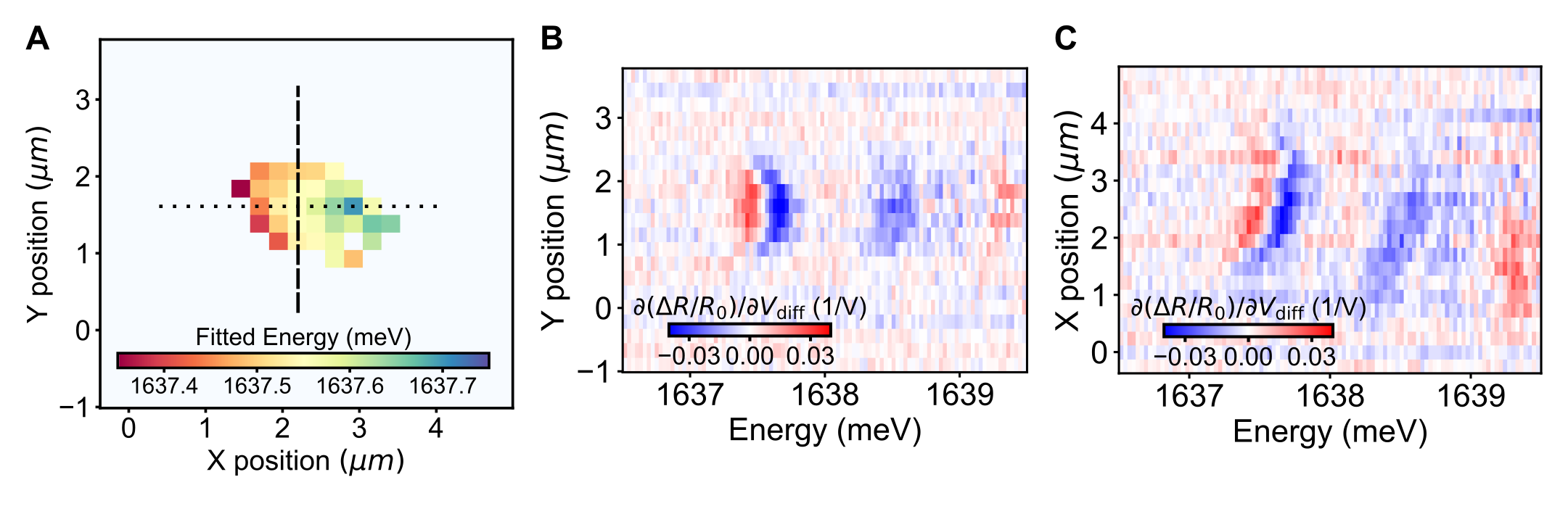}
	\caption{\textbf{Dependence of 0D exciton resonance energy on the optical excitation location.} (\bfA) Energy of the 0D resonance obtained based on the the same data and fits as in Fig.\,\ref{fig:WL1}\,{\bfF} in the main text. Only the fits with amplitude of more than $1/e$ (normalized to maximum within the dataset) are plotted for clarity. Energies of fits with an amplitude of less than this cutoff are not reliable as the resonance is either absent or too faint to be fitted properly in those regions. (\bfB) Differentiated reflectance contrast spectra measuered when the optical spot is shifted along the $y$-direction (dashed line in (\bfA)). (\bfC) Differentiated reflectance contrast spectra measuered when the optical spot is shifted along the $x$-direction (dotted line in (\bfA)).} 
	\label{fig:supp_spatial}
\end{figure*}

\begin{figure*}
	\includegraphics[width=12cm]{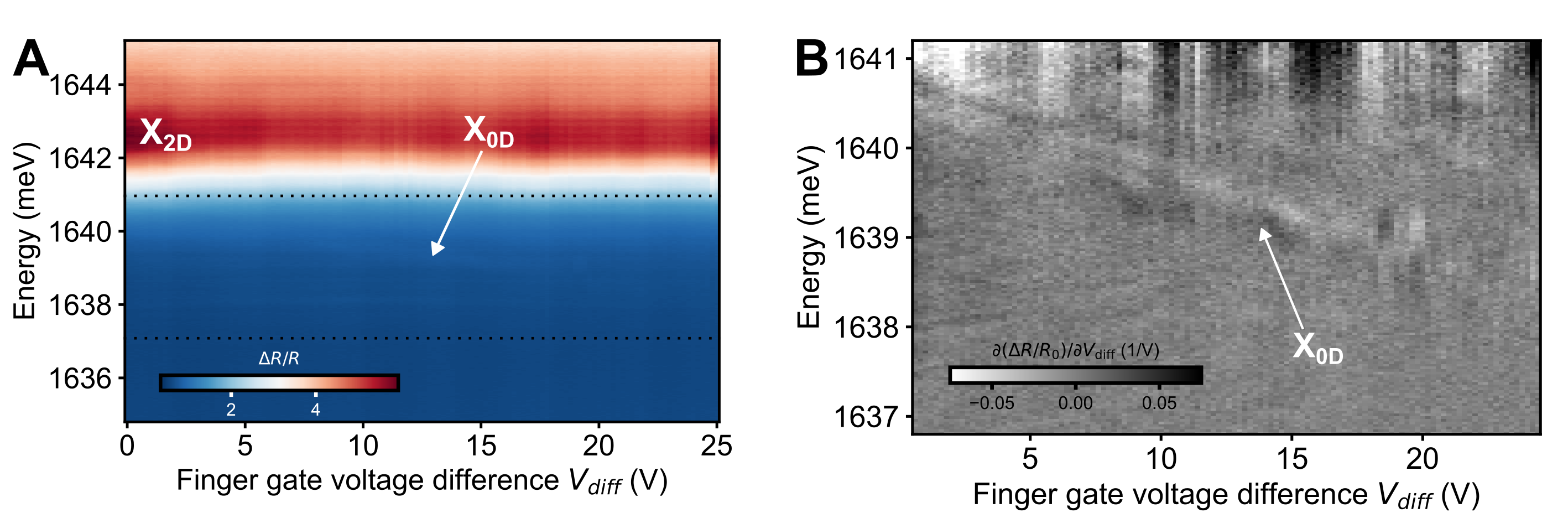}
	\caption{\textbf{Signature of 0D confined exciton in Device 2.} (\bfA) Reflectance contrast spectra for Device 2 as a function of voltage difference between finger gates. We observe a redshifting 0D resonance that is similar to that seen in Device 1, albeit with lower signal-to-noise ratio. (\bfB) Data between the dotted lines in (\bfA) differentiated with respect to applied voltage difference, which display enhanced 0D spectral feature.} 
	\label{fig:supp_dev2}
\end{figure*}

This effect is most probably responsible for a spectral shift of the 0D exciton that we observe when moving the optical spot along the finger gates direction $x$. As seen in Figs\,\ref{fig:supp_spatial}{\bfA},{\bfB}, the resonance energy decreases when the optical spot shifts towards the finger gate at lower $x$ values that hole-dopes the TMD monolayer. As the hole doping density is known to increase upon optical excitation (as demonstrated in prior experiments, e.g., ~\cite{Handa2024}), light illumination of the lower finger gate entails larger electric fields and stronger confinement, which in turn gives rise to the observed redshift of the 0D transition.

Importantly, almost no energy shift is observed when the optical spot is moved along the perpendicular $y$-direction, as shown in  Fig.\,\ref{fig:supp_spatial}{\bfC}. This observation further strengthens our claim that the observed influence of the optical spot location on the resonance energy arises exclusively due to the photo-doping effect.\\

\subsection{0D exciton confinement in Device 2}
We performed reflectance contrast measurements also in Device 2 as defined in Fig.\,\ref{fig:concept}\,{\bfB}. As shown in Fig.\,\ref{fig:supp_dev2}\, the 0D confined exciton shows up as a narrow, discrete line that redshifts as the voltage difference between the finger gates is increased, which is fully consistent with the response observed for Device 1.

\end{document}